\documentclass[10pt]{iopart}
\usepackage[utf8]{inputenc}
\usepackage[normalem]{ulem}
\usepackage{graphicx}
\usepackage{hyperref}
\usepackage{xcolor}
\usepackage{iopams}

\usepackage{mathtools}
\DeclarePairedDelimiter{\evdel}{\langle}{\rangle}
\usepackage[colorinlistoftodos]{todonotes}
\newcommand{\Pin}{P_{\mathrm{next}}}
\newcommand{\fhigh}{f_{\mathrm{high}}}
\newcommand{\flow}{f_{\mathrm{low}}}

\begin{document}

\title[Geometrically constrained folding models]{What geometrically constrained folding models can tell us about real-world protein contact maps}
\author{Nora Molkenthin$^{1*\dagger}$, J. Jasmin Güven$^{2*}$, Steffen Mühle$^{3}$, Antonia S. J. S. Mey$^{2\dagger}$}
\address{$^1$Potsdam Institute for Climate Impact Research, Germany}
\address{$^2$EaStCHEM School of Chemistry, University of Edinburgh, Edinburgh, United Kingdom}
\address{$^3$Max-Planck Institute for Dynamics and Self-Organization (MPIDS), Am Faßberg 17, 37077 Göttingen, Germany}
\address{$^*$ These authors contributed equally}
\address{$^\dagger$ To whom correspondence should be addressed}
\ead{nora.molkenthin@pik-potsdam.de,antonia.mey@ed.ac.uk}

\begin{abstract}
The mechanisms by which a protein's 3D structure can be determined based on its amino acid sequence have long been one of the key mysteries of biophysics. Often simplistic models, such as those derived from geometric constraints, capture bulk real-world 3D protein-protein properties well. One approach is using protein contact maps to better understand proteins' properties. Here, we investigate the emergent behaviour of contact maps for different geometrically constrained models and real-world protein systems. We derive an analytical approximation for the distribution of model amino acid distances, $s$, by means of a mean-field approach. This approximation is then validated for simulations using a 2D and 3D random interaction model, as well as from contact maps of real-world protein data. Using data from the RCSB Protein Data Bank (PDB) and AlphaFold~2 database, the analytical approximation is fitted to protein chain lengths of $L\approx100$, $L\approx200$, and $L\approx300$. While a universal scaling behaviour for protein chains of different lengths could not be deduced, we present evidence that the amino acid distance distributions can be attributed to geometric constraints of protein chains in bulk and amino acid sequences only play a secondary role. 

\end{abstract}
\noindent{\it Biophysics, Geometric constraints, Protein folding\/}
\maketitle
\section{Introduction}
Proteins, the molecular machines of every living organism, perform vital tasks required for life to persist. These range from transport (e.g. hemoglobin)~\cite{ahmed2020hemoglobin}, signal transduction (e.g. rhodopsin)~\cite{nagata2021rhodopsins}, immune responses (e.g. antibodies), and hormonal regulation (e.g. insulin)~\cite{dill2008protein, Dill1042}. All natural proteins are made of 20 different amino acids which dictate the 3D conformations the proteins adopt in order to function~\cite{scheraga2007proteinfolding}. One of the great challenges has been to understand how the primary structure, i.e. the amino acid sequence, can lead to the fully folded functional protein, known as the protein folding problem~\cite{nassar2021protein}. The last 50 years have seen many different routes to computationally predict biologically active protein structures without having to solve a crystal structure of the protein~\cite{marks2012protein}. AlphaFold~2, a machine learning model for structure prediction, has shown predictiveness at a level of experimental accuracy in the most recent protein structure prediction challenge~\cite{jumper2021highly, kryshtafovych2021critical}. It was trained on structural data from the Research Collaboratory for Structural Bioinformatics Protein Data Bank~\cite{berman2000protein} (RCSB PDB). The RCSB PDB contains structures from crystallographic, NMR, and cryo-EM experiments. AlphaFold~2 enabled many new avenues of research that require good protein structure prediction~\cite{callaway2022what}, but it has not solved the mechanisms by which proteins fold yet. A clear step change is that it does provide a rich resource of structures to understand emergent patterns in proteins on a more fundamental level. 

One way of mapping the functional forms of proteins and classifying their structural properties is by studying protein contact maps (PCMs)~\cite{Vendruscolo2002,dipaola2013protein,Estrada2011}. A PCM is a matrix representation of spatially close amino acids, given a certain cut-off distance as determined typically from a protein structure. PCMs have shown promise in understanding protein folding patterns, as well as revealing allosteric communication pathways~\cite{yao2019establishing, menichetti2016network,dokholyan2002topological}. Their use is also prominently featured in machine learning approaches. For example, similarly to AlphaFold~2, one challenge is to predict a PCM \textit{de novo} from sequence alone~\cite{bassot2019using, rives2021biological, rao2021msa}, without using structural data as part of the training set. Another is the use of PCMs in machine learning approaches to e.g. train a model able to predict binding affinities between a protein and a drug-like molecule~\cite{jiang2020drug}.

From a physicist's perspective, it is interesting to understand the emergent behaviour, in terms of folding or function, in the ensemble of all protein structures. The underlying features of a PCM can reveal information on the scale of the proteome rather than the individual protein and therefore, having a reliable model to generate the spatial structure hidden beneath a PCM is invaluable. As such, we consider the set of all proteins of a given length and try to find general principles to allow the characterization of this set. Effectively, this corresponds to taking an average over all proteins of a given length. This approach was already taken in~\cite{molkenthin2020self}, where it was shown that many of the simpler features of PCMs can in fact be reproduced by only considering geometric constraints of folding chains, disregarding all additional complexities arising from the specific features of a sequence or even secondary structures.

Different heuristic models for such simplified PCMs have been introduced in the past by Atilgan \textit{et al.}~\cite{atilgan2004smallworld} and Bartoli \textit{et al.}~\cite{bartoli2008effecta}. Atilgan \textit{et al.} propose a random shuffle model to generate artificial PCMs with similar properties to real protein models. Bartoli \textit{et al.} built a model for PCMs in which they assume ~\textit{ad hoc} that proteins have a distribution of amino acid distances $P(s)$ that follows $P(s) \approx s^{-1}$. The \textit{amino acid distance} $s$ is the separation of two connected amino acids along the backbone chain and gives rise to the simplest implementation of a connection probability in a protein. Bartoli \textit{et al.} justify their approximation only heuristically in the sense that resulting PCMs have similar properties to real-world PCMs. 

In this paper, we provide an explanation for why the $s^{-1}$ heuristic is a good initial assumption. To do so, we derived an analytical approximation from a 2D geometrically constrained model to get an expression for $P(s)$. The approximation provides additional correction terms. We then show that the approximation can be used to fit simulations of the 2D constrained model~\cite{molkenthin2016scaling}, a 3D version of the constrained model~\cite{molkenthin2020self}, and even to real-world proteins. This means that the analytical approximation is a good starting point for artificially modelling PCMs in the future. 


The paper is structured in the following way: In section~\ref{sec:models}, we give a brief overview of the geometrically constrained models used. In section~\ref{sec:methods}, we describe the construction of PCMs from the 2D and 3D simulations of the geometrically constrained model, as well as from structural data in the RCSB PDB and AlphaFold2 databases. In section~\ref{sec:theory} we then derive an analytical approximation for the 2D analog of the geometrically constrained model. Section~\ref{sec:results} highlights the results for the 2D, and 3D simulations, RCSB PDB, and AlphaFold~2 structural data by making use of the 2D analytical approximation.
\section{2D and 3D geometrically constrained models}
\label{sec:models}
Different approaches have been used in the past for geometrical models describing protein folding, some of which derived characteristics of the secondary structure from constraints on bond and torsion angles~\cite{bhattacharjee2013flory, Danielsson2010, Molkenthin2011} or on the formation mechanisms of the tertiary structure~\cite{molkenthin2016scaling, molkenthin2020self}, and others are modeled by self-avoiding random walks~\cite{mey2014rareevent, hills2009insights}. Here we will focus on models representing maximally folded structures in 2D, and 3D as introduced in~\cite{molkenthin2016scaling} and~\cite{molkenthin2020self}, respectively. The 3D geometrical model and the analytical approximation of its 2D analogue both build on the idea that inherently geometrical objects, such as amino acids, imply that any resulting PCM network ensemble modelling them has to be spatially embedded. 

\begin{figure}[h!]
    \centering
    \includegraphics[width=0.5\columnwidth]{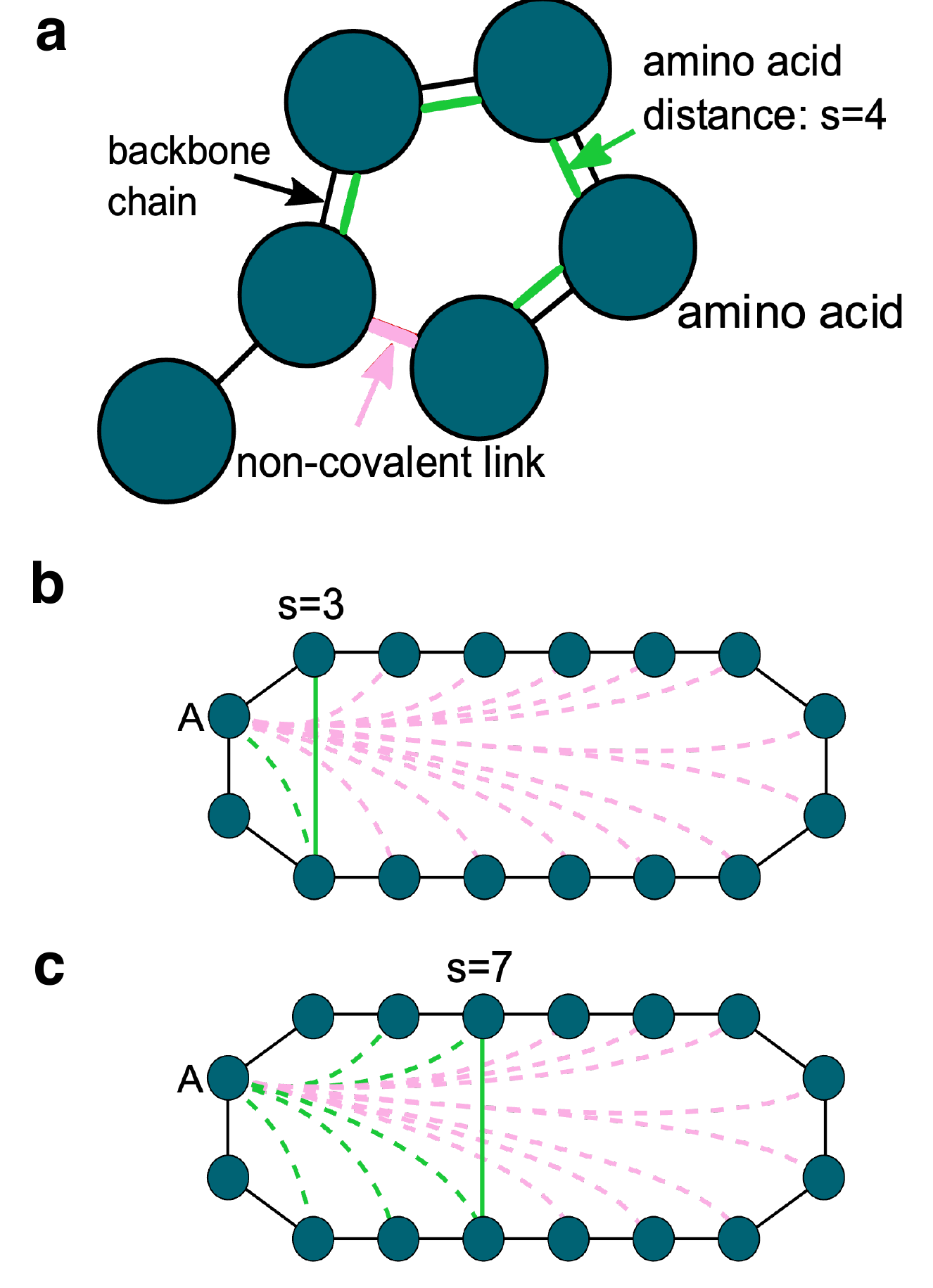}
    \caption{a) The amino acid distance $s$ (green) is defined as the number of amino acids (AAs) (here represented by a circle) along the protein backbone chain (black) between two connected AAs (pink). b) Shows an example AA distance of length $s=3$ (solid green line). When adding a new connection from node A, the previously created connection prohibits new ones (dashed pink lines). The green dotted link from A is still permitted and can be made. c) Shows an example of an existing AS distance $s=7$ (solid green), where there are now more permitted (dotted green) links and fewer prohibited (dotted pink) ones.}
    \label{fig:schematic}
\end{figure}

In~\cite{molkenthin2016scaling}, we introduced a simplified, 2D version of this geometrically constrained model. It starts from a closed chain (i.e. a ring) of $L$ unit discs and subsequently adds links, such that connected discs touch, yet no discs intersect. The advantage of this model is that it can be approximated by a purely topological simplification, which can be treated analytically. Figure~\ref{fig:schematic} illustrates the construction of the 2D geometric model. The 3D version gives rise to the geometric constraints directly, rather than through an approximation by means of a topological constraint.

The topological constraints in the resulting network model are:
(a) New links always form between two units that are part of the same face of the graph (region enclosed by a cycle in the network). This prevents the overlapping of discs. (b) No links form across the outer face. This prevents the enclosure of a unit by less than six other units (which is geometrically impossible) such that (c) the maximum degree of each unit is six, as six is the maximum number of unit discs one central unit disc can touch. (d) Once connected by a link, pairs of units do not disconnect. In~\cite{molkenthin2020self} this geometrically constrained model was extended to 3D spheres using simulations to generate compact artificial protein-like polymer structures. 
\section{Protein contact maps (PCM)}
\label{sec:methods}
The complex interaction pattern between the amino acids in a protein can be naturally expressed as a network or graph, in which each amino acid is represented by a node and spatial proximity is encoded as a link. Whenever two central $\mathrm{C}_\alpha$ atoms are closer together than a threshold $d_c$, they are connected, linked, or in `close contact'. These connections can be determined from the 3D functional or folded structure of the protein and are encoded in a so-called (protein) contact map. This contact map is effectively an adjacency matrix $\textbf{A}^{\mathrm{struc}}$ 
\begin{equation}
  A^{\mathrm{struc}}_{ij}=
  \begin{cases}
   0, & \text{ if } d_{i,j}>d_c \text{ or } i=j\\
      1, & \text{ if } d_{i,j}\leq d_c.
      \end{cases}
    \label{eq:aij}
\end{equation}

\subsection{Contact maps in a geometrically constrained protein model}

Based on the simple observation that amino acids are objects in space that cannot overlap indefinitely, in~\cite{molkenthin2020self} we introduced the 3D version of the 2D geometrically constrained model. Starting from a chain of identical spheres, each in contact only with the neighbours it is connected to, additional connections are made by randomly selecting two spheres and moving them towards each other until they touch. The new links formed that way cannot be broken in later steps and function as constraints on subsequent link formation steps. If the contact of the two selected spheres is geometrically impossible without breaking previously made connections or leading to overlaps of any spheres, the link is not made and is taken out of the pool of possible connections. This process is repeated until no more links can be formed without violating the geometric constraints and the final structure can be expressed as the adjacency matrix
\begin{equation}
  A^{\mathrm{sim}}_{ij}=
  \begin{cases}
   1, & \text{ if }|i-j|=1 \text{ or i and j connected} \\
      0, & \text{ otherwise}.
      \end{cases}
    \label{eq:sim_aij}
\end{equation}



\subsection{Protein contact maps from structural protein data}

\begin{figure}[htb]
        \centering
	\includegraphics[width=0.5\columnwidth]{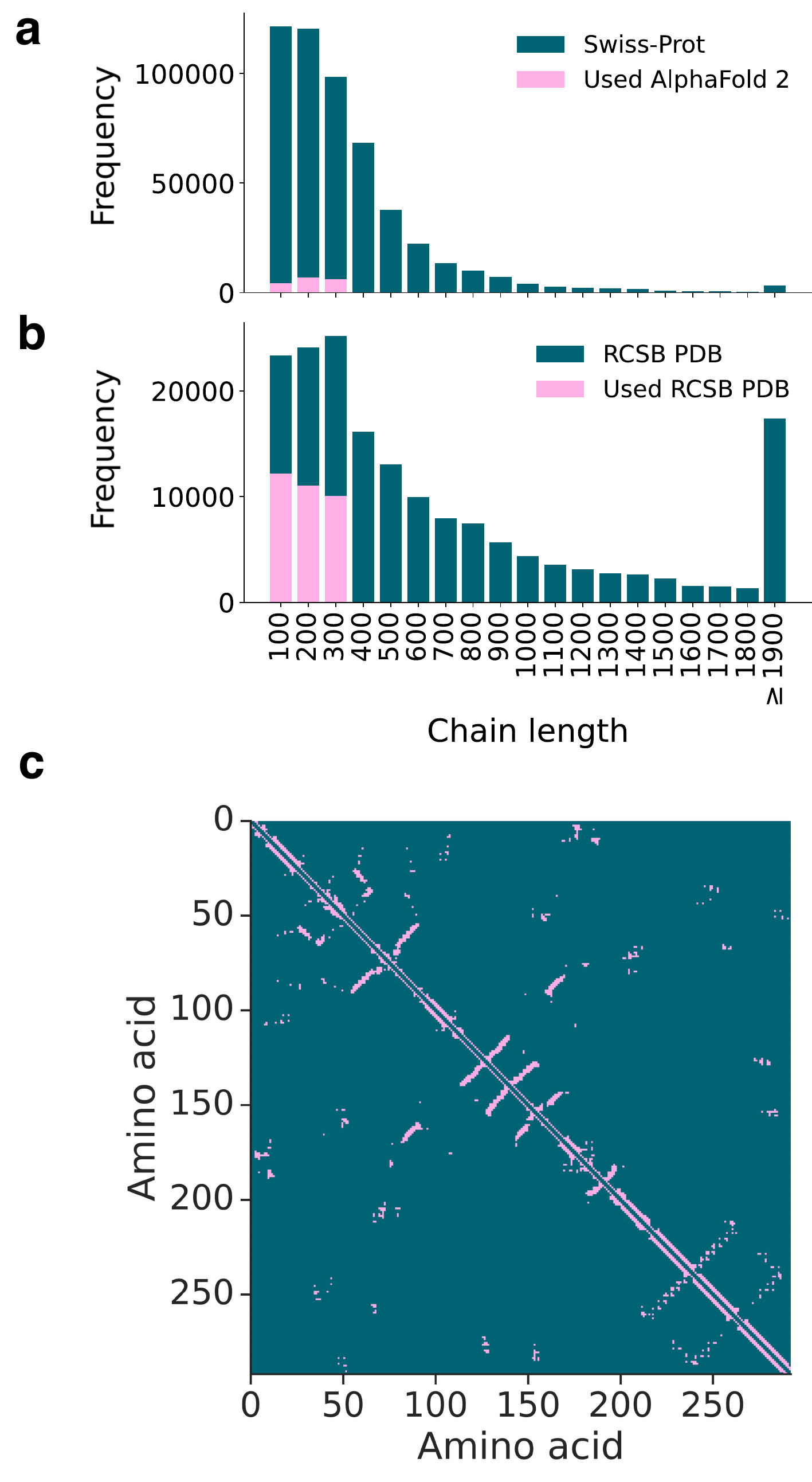}
	    \caption{Bar plot of protein sequence lengths for protein structures from two protein structure databases. a) AlphaFold~2 predicted protein structures taken from protein sequences in the manually curated SwissProt database of protein sequences. Pink bars show the fraction of structures used for the average contact map analysis. b) RSCB PDB chain length frequencies. Pink bars show protein structures whose chain lengths match the resolution of the structure accounting for the whole sequence. Only pink RCSB PDB samples were used for the average contact map analysis. c) Example of a protein contact map taken from RCSB PDB ID: 3UFG, with a cut-off of 8~Å. Pink squares indicate a contact present between amino acids.}
        \label{fig:pdb_stats}
\end{figure}

The underlying formation mechanisms of individual protein folds are incredibly complex and depend on the surrounding solvent, as well as the specific amino acid sequence and their interactions. Here, we assess if the geometrically constrained protein model's contact map distributions capture real protein behaviour and make it a viable model for PCMs and provide a better explanation for the heuristic used by Bartoli \textit{et al.}~\cite{bartoli2008effecta}. We study the ensemble of PCMs from real protein structures and will look at `averaged' contact maps over many different proteins that have different amino acid sequences, but their overall sequence or chain length is the same. 
We use the RCSB Protein Data Bank (RCSB PDB), a database of structurally resolved protein amino acid sequences through X-ray crystallography, NMR or cryo-EM experiments and the synthetically generated structural database from the machine-learned structures from the AlphaFold~2 database~\cite{varadi2022alphafold}. The RCSB PDB contains just under 200,000 protein structures to date~\cite{berman2000protein}. The distribution of protein chain lengths in the RCSB PDB is illustrated in Figure.~\ref{fig:pdb_stats} b. These chain lengths vary from small fragments of less than 10 amino acids to large agglomerates of over 2000 amino acids. The most frequently occurring chain lengths, however, are between 100 and 300 amino acids as shown in figure~\ref{fig:pdb_stats} b. The pink subset of the bar plot shown in figure~\ref{fig:pdb_stats} b is the set of structures used that fulfil the criterion on protein chain lengths of $L\approx100$ ($85\le L\le 115$) amino acids, $L\approx 200$ ($185 \le L \le 215$) amino acids and $L\approx300$ ($285 \le L\le 315$) amino acids and represent all RCSB PDB structures used in our analysis. The second database consists of synthetically generated structures using AlphaFold~2 with sequences taken from SwissProt~\cite{bairoch2000swissprot}. SwissProt is a manually curated database of protein sequences containing over 500,000 protein sequences, whose protein chain length distribution is shown in figure~\ref{fig:pdb_stats} in teal. The pink bars indicate the subset of AlphaFold~2 structures used for the contact map analysis. In this case, the initial structures were selected based on the condition of chain lengths in the same three intervals as used for the RCSB PDB. Second, the AlphaFold~2 per-residue confidence scores~\cite{jumper2021highly} were used to select proteins that had an average score of 90 or higher. Lastly, we removed structures that described the same protein but originated from different organisms, by selecting only unique protein names. 

All structure-based PCMs were generated in the same way for both AlphaFold~2 structures and RCSB PDB structures. Structures used were downloaded in pdb format, read in with MDAnalysis version 2.0.0~\cite{gowers2016mdanalysis} and only C$_{\alpha}$ were selected and used for the analysis. Proteins were categorised by length and placed into the same three groups as for the RCSB PDB data. For each of the sequences, all pairwise C$_\alpha$ distances were calculated, and two atoms were said to be in contact if their distance was less than 8~Å. The contacts (1 or 0) were then recorded in the adjacency matrix $\mathbf{A}^{\mathrm{struc}}$. An example of a protein contact map is shown in figure~\ref{fig:pdb_stats}~c. We study the distribution of distances between connected amino acids in real and simulated proteins calculated from adjacency matrices. These distances were then histogrammed to calculate the mean amino acid distance distributions shown in section~\ref{sec:results}. The pseudo-algorithms describing the above two processes are shown in the Supplemental Material~(SI)~\cite{SI}. All code for the construction of the contact maps and their analysis is available on GitHub at~\cite{2022sequence}. The derived analytical approximation in section~\ref{sec:theory}, was used for data fitting. All fits were carried out using SciPy 1.7.0 and with scripts for the fits available on GitHub at~\cite{2022sequence}. The secondary structure protein analysis was done using a local DSSP server~\cite{1983kabsch}, and the scripts for the processing of the DSSP output are available in the same GitHub repository~\cite{2022sequence}.

\section{Analytical approximation for the geometrically constrained model}\label{sec:theory}

To assess how we can best describe the distribution of the probability of amino acid distances $P(s)$ obtained from an average PCM for the geometrically constrained model, we consider the amino acid distance distribution for the simplified 2D model as proposed in~\cite{molkenthin2016scaling}. We make the assumption that each distance is equally likely, from which we propose an analytical approximation for $P(s)$, derived from the geometric models, and schematically summarised in figure~\ref{fig:schematic}.

We introduce the auxiliary variable $F_k(s)$, defined to be the number of possible links with an amino acid distance of $s$ if $k$ links have been added before. Before any links are added, because we are looking for an average PCM over the whole proteome restricted to a particular chain length, we make the assumption that all amino acid distances are equally likely, so there are $F_0(s)=N$ possibilities of making a link of distance $s$ each, with $2\leq s < \frac{N}{2}$.

As we add more links, not only are links taken out of this pool, because they have already been realized, but each existing link can also geometrically prohibit other connections. Adding a link of length $s_1$ prohibits $2(s-1)$ links of length $s<s_1$ and $2(s_1-1)$ links of length $s\geq s_1$ (see figure~\ref{fig:schematic}~b). To find the expected number of links still available in the second step, we average over all possible amino acid distances $s_1$ of the first link added.

The expected distribution of possible links after one step is thus given by the average over available link pools taken over all possible values of $s_1$:

\begin{align}
    F_1(s)&=\frac{1}{\frac{N}{2}-1} \sum_{s_1=2}^{N/2}{ \begin{cases}
    F_0(s)-2(s-1) \text{ , for } s<s_1\\
    F_0(s)-2(s_1 -1)\text{ , for } s\geq s_1
    \end{cases}}\nonumber\\
    &\approx\frac{2}{N} \sum_{s_1=2}^{N/2} { \begin{cases}
    N-2(s-1) \text{ , for } s<s_1\\
    N-2(s_1 -1)\text{ , for } s\geq s_1.
    \end{cases}}
\label{eq:reduction_first_step}
\end{align}

After $k-1$ links are made, the probability that the $k^{\mathrm{th}}$ link, being randomly chosen from the pool of available links, has length $s$ obeys
\begin{equation}
    \Pin^k(s)=\frac{F_k(s)}{C_k},
    \label{eq.Pnext}
\end{equation}
where $C_k=\sum_{s=2}^{N/2}F_k(s)$. In subsequent steps, we use this probability to perform a weighted average over new links being added and get the same reduction (compare figure~\ref{fig:schematic}) but starting from the pool $F_{k-1}(s)$ of the step before rather than the initially available $N$ links. 
The probability of a link of length $s$ is thus the average overall added link probabilities from the first to the last added link. In 2D, this is repeated for $N-3$ steps until no more links can be added \cite{molkenthin2016scaling}, giving
\begin{equation}
    P(s)=\frac{1}{N-3}\sum_{k=0}^{N-4} \Pin^k(s).
    \label{eq.P_next_sum}
\end{equation}
However, we must subtract fewer links to account for some of them having left the pool in earlier steps. To this end, we multiply the number of blocked links by
\begin{equation}
    \frac{F_{k-1}(s)}{F_0(s)} = F_{k-1}(s)\frac{1}{N}.
\end{equation}

This ignores any length-dependent bias in which links have been removed. Having considered some examples this assumption seems to hold well. 
This leads to the following recursion:

\begin{align}
    F_k(s)&= \sum_{s_k=2}^{N/2} \Pin^k(s_k)\nonumber \bigg(F_{k-1}(s) - F_{k-1}(s)\frac{1}{N}
    {\begin{cases}
     2(s-1) \text{ , for } s<s_k\\
     2(s_k -1))\text{ , for } s\geq s_k
    \end{cases}}\bigg),
\end{align}

which generalizes~\ref{eq:reduction_first_step} and simplifies to

\begin{align}
   F_k(s)&=F_{k-1}(s)\Big(1-\frac{1}{N} \sum_{s_k=2}^{s} 2 (s_k-1) \Pin^k(s_k) -\frac{1}{N} \sum_{s_k=s+1}^{N/2} 2 (s-1) \Pin^k(s_k)\Big) \nonumber \\
   &=F_{k-1}(s)\bigg(1-\frac{1}{N}
   \sum_{s_k=2}^{s} 2 (s_k-1) \Pin^k(s_k)-\frac{2 (s-1)}{N} \Big[\sum_{s_k=2}^{N/2} \Pin^k(s_k) - \sum_{s_k=2}^{s} \Pin^k(s_k)\Big]\bigg) \nonumber \\
   \nonumber
\end{align}
and thus:
\begin{equation}
   F_k(s) =  F_{k-1}(s)\Big(1-\frac{2}{N} (s-1) 
   +\frac{2}{N} s
   \sum_{s_k=2}^{s}\Pin^k(s_k) -\frac{2}{N}
   \sum_{s_k=2}^{s} s_k \Pin^k(s_k)\Big).\label{eq.Fk_rec}
\end{equation}

Together with the initial condition $F_0(s)=N$, this expression constitutes an iteration rule from which $F_k(s)$ can be found for all $k$ and $s$. Aiming to decouple those dynamics for different $s$, we make a heuristic guess $\tilde{P}_{\text{next}}^k(s_k)$ for $\Pin^k(s_k)$. It consists of two separate approximations which are used in steps $k<a$ and steps $k\geq a$ respectively, where the threshold $1\leq a\leq \frac{N}{2}$ is a free parameter. In the early stage, we use a uniform probability
\begin{equation}
    \tilde{P}_{\text{next}}^{k\ll N/2}(s)\approx \Pin^{k=0}(s)=\frac{2}{N}.
    \label{eq.P_small}
\end{equation}
For later steps, the probability of a link to still being in the pool decreases with $s$. We thus approximate $\Pin(s)$ with an ansatz for the $k-$average, namely, that it drops off as $s^{-1}$, leading to
\begin{equation}
    \tilde{P}_{\text{next}}^{k\gg 1}(s)\approx \sum_{k=a}^{N-4}\Pin^{k}(s)
    \approx \frac{s^{-1}}{ H_{\frac{N}{2}}-1},
    \label{eq.P_large}
\end{equation}
where $H_{\frac{N}{2}}$ is the harmonic number serving as a normalization factor. 
As we will see later, this approximation holds most precisely for intermediate values of $k$, with the $k-$average of $\tilde{P}_{\text{next}}^{k\gg 1}(s)\approx P(s)$, making it self-consistent. Errors for smaller and larger values of $k$ are thought to approximately cancel out.
Substituting~\ref{eq.P_small} and~\ref{eq.P_large} into~\ref{eq.Fk_rec} results in the following two recursions for the early and late evolution of the amino acid distance distribution respectively.
\begin{align}
   F_{k\ll \frac{N}{2}}(s)&\approx 
   F_{k-1}(s)\Big(1-\frac{2}{N} (s-1) 
   +\frac{4}{N^2} s (s-1)\nonumber -\frac{2}{N^2}(s^2+s-2)\Big)\nonumber \\
   &=F_{k-1}(s)(1-\flow(s)),
   \label{eq.Fk_rec_low}
\end{align}
where 
\[\flow(s)=-\frac{2}{N^2}s^2+\left(\frac{2}{N}+\frac{6}{N^2}\right)s-\left(\frac{2}{N}+\frac{4}{N^2}\right),\]

and

\begin{align}
   F_{k\gg1}(s)&\approx F_{k-1}(s)\Big(1-\frac{2}{N} (s-1) \nonumber+\frac{2}{N} s \sum_{s_k=2}^{s}\frac{1}{s_k(H_{\frac{N}{2}}-1)}\nonumber-\frac{2}{N} \sum_{s_k=2}^{s} s_k \frac{1}{s_k(H_{\frac{N}{2}}-1)}\Big)\nonumber \\
   &=F_{k-1}(s)\bigg[1 -\Big(\frac{2}{N}-\frac{2}{N}\frac{H_{s}-1}{H_{\frac{N}{2}}-1} \nonumber +\frac{2}{N(H_{\frac{N}{2}}-1)}\Big)s \nonumber + \frac{2}{N(H_{\frac{N}{2}}-1)} + \frac{2}{N}\bigg]\nonumber \\
   &=F_{k-1}(s)(1-\fhigh(s)),
   \label{eq.Fk_rec_high}
\end{align}
where 
\[\fhigh(s)=\frac{2}{N}\left(\frac{H_{\frac{N}{2}}-H_{s}+1}{H_{\frac{N}{2}}-1}s-\frac{H_{\frac{N}{2}}}{H_{\frac{N}{2}}-1}\right).\]

We can now use the recursive expressions~\ref{eq.Fk_rec_low} and~\ref{eq.Fk_rec_high} to write down a closed expression for $F_k(s)$, using $\flow$ for the first $a$ steps and $\fhigh$ for the rest:

\begin{align}
    F_k(s)&=N\prod_{i_1}^a\left(1-\flow(s)\right)\prod_{i=a+1}^k\left(1-\fhigh(s)\right) \nonumber \\
    &= N\left(\frac{1-\flow(s)}{1-\fhigh(s)}\right)^a\left(1-\fhigh(s)\right)^{k}.
    \label{eq.F_final}
\end{align}

\begin{figure*}[htb]
        \centering
	\includegraphics[width=\textwidth]{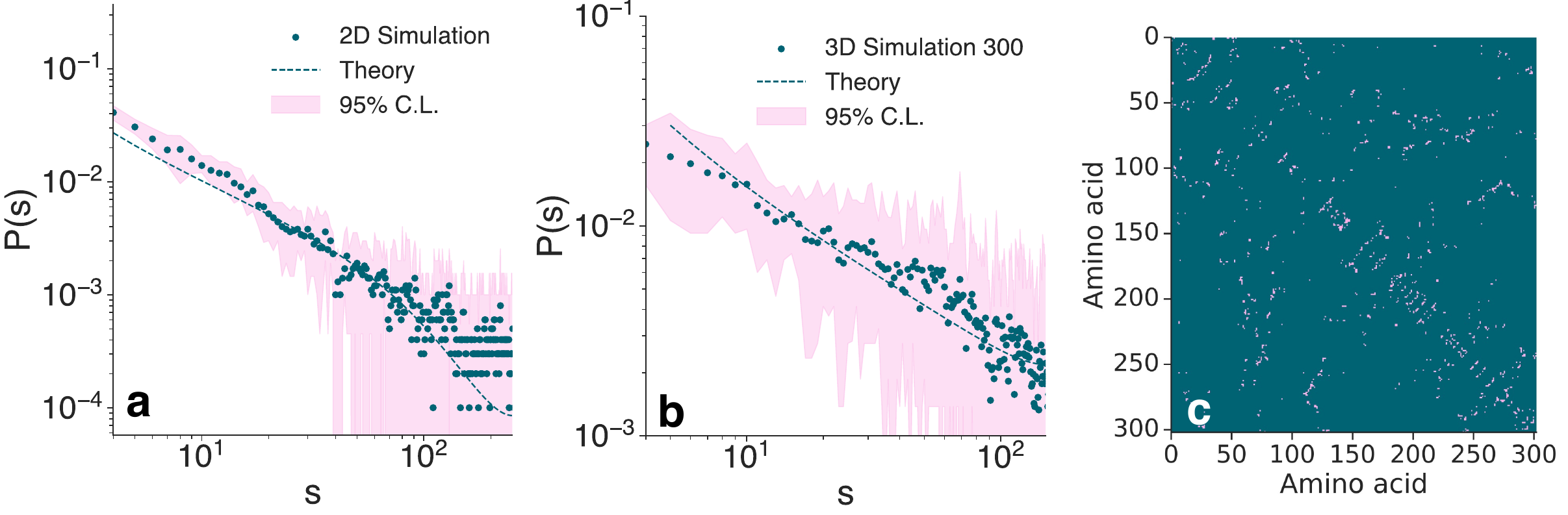}
	    \caption{The analytical approximation fitted to amino acid distance distributions for a) 2D and b) 3D simulated data. The pink area shows the 95\% confidence interval. c) Example of a PCM taken from one of the 3D simulation runs. The pink squares indicate a contact between two nodes.}
        \label{fig:2d_sim}
\end{figure*}

This can now be used to find an approximation for the amino acid distance distribution $P(s)$. By inserting~\ref{eq.F_final} into~\ref{eq.P_next_sum}, using~\ref{eq.Pnext}, we can state that the probability distribution of the realized amino acid distances of all added links is then given by the average over the available pools at each link addition step.

\begin{align}
    P(s)&=\frac{1}{N-3}\sum_{k=0}^{N-4} \Pin^k(s)  \\
    &\approx \frac{N}{N-3}\left(\frac{1-\flow(s)}{1-\fhigh(s)}\right)^a 
\times \sum_{k=0}^{N-4} \frac{1}{\evdel{C_k}_k}\nonumber \left(1-\fhigh(s)\right)^{k} 
\end{align}
where $\evdel{C_k}_k$ approximates the individual $C_k$'s as the average of $C_k$ over $k$.
This approximation allows the use of the geometric series for solving the equation:
\begin{align}
    P(s)&\approx \left(\frac{1-\flow(s)}{1-\fhigh(s)}\right)^a\frac{\Gamma}{\fhigh(s)},
    \label{eq.Ps}
\end{align}
where $\Gamma=\frac{N}{(N-1)}\evdel{C_k}_k$ collects all constant factors and is used as a free fitting parameter, as the details of the evolution of $C_k$, as well as the role of dimensionality (3D vs. 2D) are unknown.
The resulting expression in~\ref{eq.Ps} by visual inspection of figure~\ref{fig:2d_sim} resembles a power law, thus justifying the earlier approximation of $\Pin^k(s)\approx\tilde{P}_{\text{next}}^k(s)$ introduced in~\ref{eq.P_large}.

\section{Results and Discussion}\label{sec:results}
Our analysis centres around understanding how the analytical approximation can be used to understand the simulations in 2D and 3D from~\cite{molkenthin2016scaling,molkenthin2020self}, and structural data from the RCSB PDB and AlphaFold~2 databases. To this end,~\ref{eq.Ps} is used as a fitting function in order to determine the two parameters $\Gamma$ and $a$.

\subsection{The analytical approximation fits 2D and 3D simulation data}
Figure~\ref{fig:2d_sim} compares~\ref{eq.Ps} with respect to (a) 2D and (b) 3D simulation data introduced in~\cite{molkenthin2016scaling} and ~\cite{molkenthin2020self}, respectively. Figure~\ref{fig:2d_sim}~c shows an example of an adjacency matrix $\mathbf{A}^{\mathrm{sim}}$ as generated from a 3D simulation run. The 2D simulation data was generated from 10 repeats with chain lengths of 498 amino acids taken from~\cite{molkenthin2016scaling}. The 3D simulation data consists of adjacency matrices computed from 30 simulation runs and was analyzed as was explained in section~\ref{sec:methods}. The amino acid distance distributions are plotted to show the mean frequencies of each amino acid distance up to $\frac{L}{2}\sim250$ and $\frac{L}{2}\sim150$ for 2D and 3D simulations, respectively. The shaded areas represent the 95\% confidence intervals. The theoretical approximation from~\ref{eq.Ps} was fitted to both 2D and 3D simulation data separately. The two parameters, $a$~and~$\Gamma$ in~\ref{eq.Ps}, are given in Table~\ref{table:sim-results}. See the SI~\cite{SI} for the parameters for 3D simulation data in the $L\approx 100$ and $L \approx 200$ ranges. The fitted approximation captures the behaviour of the simulation data well for both~2D and~3D simulations.

\begin{table}[htb]
\centering
\setlength{\tabcolsep}{25pt}
\begin{tabular}{ccc}
\hline
Simulation & a &$\Gamma\; (10^{-3})$ \\
\hline
2D& $6 \pm 1$ & $0.3 \pm 0.1$ \\
3D 300& $1 \pm 1$ & $7.0 \pm 0.2$ \\
\hline
\end{tabular}%
\caption{Fit parameters from fitting the analytical approximation to the 2D simulated data as well as the 3D simulated data in the $L\approx 300$ chain length range.}
\label{table:sim-results}
\end{table}

\subsection{$P(s)$ from RCSB PDB and AlphaFold~2 agree and cover similar secondary structure content}
Since AlphaFold~2 generates structures from an AI model, the database contains a variety of structures whose 3D folded structures are more or less confident (as measured by the confidence score measure provided by the database). Therefore, before analyzing the amino acid distance distributions from RCSB PDB and AlphaFold~2, we wanted to understand if the distributions generated from the two databases at lengths $L\approx 100$, $L \approx 200$, and $L \approx 300$ are statistically the same, given the selection of a mean confidence score of 90 or above.
To this end, we used the non-parametric test of the Kolmogorov-Smirnov (KS) statistic to compare the probability distributions of AlphaFold~2 and RCSB PDB for each of the lengths~\cite{2008kolmogorov}. The KS statistic for the lengths of $L \approx 100$, $L \approx 200$, and $L \approx 300$ amino acids was 0.07, 0.10, and 0.12 respectively. Therefore, the null hypotheses for the distributions being the same were accepted at the $\alpha=0.01$ level. 

In addition to calculating the KS statistic between the filtered AlphaFold~2 structures (confidence score above 90) and RCSB PDB structures, we also used a subset of AlphaFold~2 structures in the $L \approx 300$ range with per-residue confidence scores below 70. In this case, the KS statistic was 0.50 and the null hypothesis was rejected at the $\alpha=0.01$ level. For more details see the SI~\cite{SI}. 

The number of structures used for the analysis in each chain length range for both databases is shown in Table~\ref{table:subsets}. All structures were converted into PCMs, as described in Section~\ref{sec:methods}. From the PCMs we computed the amino acid distance distribution in the same way as was done for the 3D simulations.

\begin{table}[htb]
\centering
\setlength{\tabcolsep}{5pt}
\resizebox{0.5\columnwidth}{!}{
\begin{tabular}{ccc}
\hline
Protein chain length $L$ & RCSB PDB & AlphaFold 2 \\
\hline
100 & 12,206 & 4,396 \\
200 & 11,031 & 7,053 \\
300 & 10,091 & 6,231 \\
\hline
\end{tabular}
}
 \caption{The number of structures used for the analysis from the RCSB PDB and AlphaFold 2 database, after filtering.}
\label{table:subsets}
\end{table}

\subsection{The analytical approximation can describe the amino acid distance distributions of the structural data}
Next, we investigated if the amino acid distance distribution from the structural data (RCSB PDB and AlphaFold~2) can be modeled by the analytical approximation.
In order to answer this, we look at the scaling of the distribution of RCSB PDB data in the first instance.

\begin{figure*}[htb]
        \centering
	\includegraphics[width=\textwidth]{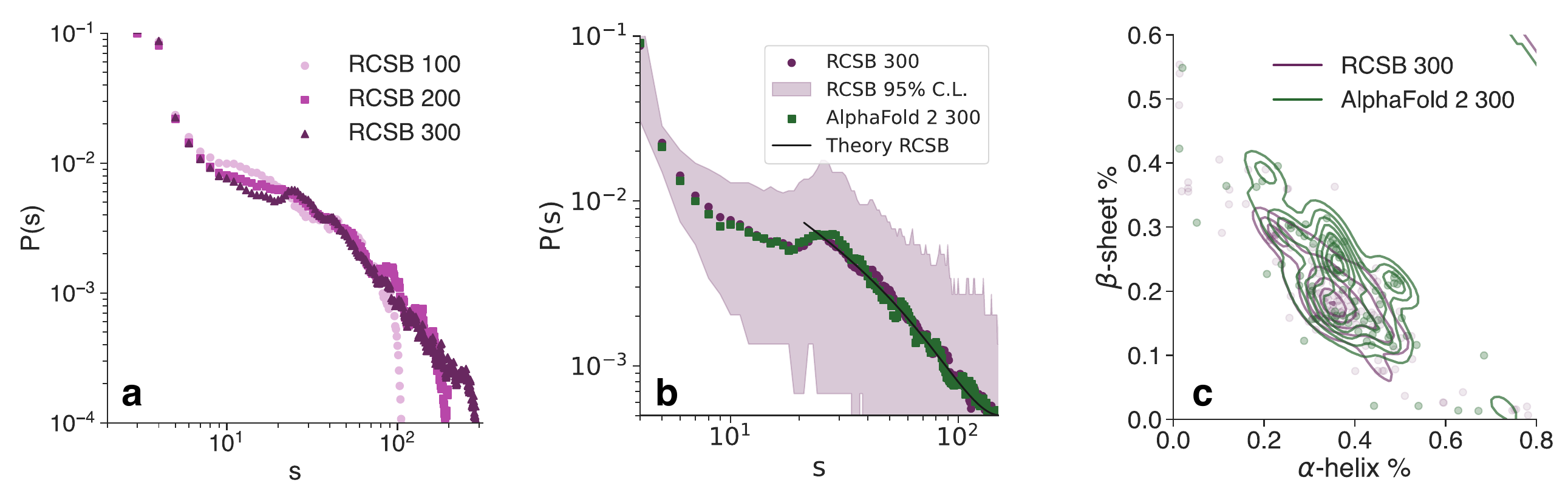}
        \caption{a) Amino acid distance distributions for contact maps derived from the RCSB PDB database for protein chains in the ranges of $L \approx 100$, $L\approx200$, and $L\approx300$ amino acids. b) Comparison between the distributions of RCSB PDB and AlphaFold~2 database for protein chains in the $L \approx 300$ range. The analytical approximation (solid line) is fitted to the tail of the RCSB data. The pink shaded area shows a 95\% confidence level. c) Secondary structure content distribution in the RCSB PDB and AlphaFold~2 structures in the $L\approx300$ chain length range.}
        \label{fig:sdd}
\end{figure*}

Figure~\ref{fig:sdd} a shows the distributions of the amino acid distances for RCSB PDB logarithmically scaled for the different protein chain lengths of $L\approx100$ (light purple circles), $L\approx200$ (medium purple squares), and $L\approx300$ (palatinate triangles). For all three lengths, the distributions show an approximate power law decay in their tail, which is consistent with the simulated results. 

We fit the analytical approximation to the RCSB PDB amino acid distance distribution in figure~\ref{fig:sdd} b in the $L\approx300$ range. In addition, we show the AlphaFold~2 amino acid distance distribution for comparison. The theoretical fit is represented by a solid black line. The parameters given by the analytical approximation are $a=4 \pm 1$ and $\Gamma=(7.4 \pm 0.8)\cdot 10^{-4}$. The shaded area shows a 95\% confidence level (C.L.). Similar fits for $L\approx100$ and $L\approx200$ as well as the AlfaFold~2 results are presented in the SI.

We observe that for intermediate values of $s$, i.e. $30\lesssim s \lesssim \frac{L}{2}$, the distributions are described well by the analytical approximation. However, the curves deviate for amino acid distances in the $5\lesssim s \lesssim 30$ range (see figure~\ref{fig:sdd} b). Comparing this to the simulation distributions and their theoretical fits (Figure~\ref{fig:2d_sim}), we see an under-representation of amino acid distances calculated from PCMs from RCSB PDB data. 

To understand this under-representation better, we looked at the secondary structure content of the PCMs we studied. The reasoning behind this is that the range in which the under-representation is observed is at a specific distance range, where contacts arise from the proteins' secondary structure elements. In figure~\ref{fig:sdd}~c we show the secondary structure content distributions for proteins in the $L\approx300$ chain length range from RCSB PDB and AlphaFold~2. We see that proteins mostly contain secondary structures in the combined $\alpha$-helix---$\beta$-sheet region. For protein chain length regions of $L\approx100$ and $L\approx200$ proteins with predominantly $\alpha$ and $\beta$-sheet regions exist, are shown in the SI~\cite{SI}. However, in all data, a high percentage of mixed secondary structure is observed. This is also in agreement with~\cite{michie1996analysis}. As all three chain length ranges contain high percentages of both mixed and predominantly $\alpha$ or predominantly $\beta$ secondary structures. This rules out that a certain type of secondary structure, or absence thereof, gives rise to this under-representation the analysis does not give insights if the secondary structure, in the form of an $\alpha$-helix---$\beta$-sheet as a structure per se leads to this under-representation.

Threfore we next looked at different threshold used that define a contact between two amino acids. We changed the threshold value $d_c$ from the original $8 \; \mathrm{\AA}$ to 10, 15 and 21 $\mathrm{\AA}$, see figure 4 in the SI~\cite{SI}. For 10 $\mathrm{\AA}$, the shape of the distribution was similar to that of in figure~\ref{fig:sdd}~b, but for 15 $\mathrm{\AA}$ the under-representation starts to disappear, whereas for 21 $\mathrm{\AA}$ it disappears almost completely. It is thus likely that with a threshold of 8 $\mathrm{\AA}$ some inter-secondary structure contacts are being excluded, e.g. for large $\beta$-sheet content. Tuning this threshold value allows for the modulation of the under-representation of amino acid distances in the $5 \lesssim s \lesssim 30$ region. Compared to simulations where this is not observed due to the neglect of the existence of secondary structure.

\section{Conclusion}
The tertiary structures of folded proteins have long been one of the most important mysteries of biophysics. While for individual structures, detailed molecular dynamics or statistical models are essential in approaching these questions, general statements for the ensemble of folded proteins can be made with much simpler models.

Here, we introduced and analyzed the amino acid distance $s$ and its probability distribution $P(s)$ for both real-world and gemoetrically constrained model simulations. For the equivalent model in 2D, we have used a mean-field approach to derive an analytical approximation for the amino acid distance distribution $P(s)$, and we show its agreement with the simulated geometrically constrained folding model and measured distributions from real-world proteins.
Therefore we demonstrate that the geometrically constrained model's amino acid distance distribution $P(s)$ can match real-world data well. In addition, the derived analytical approximation can serve as a good basis to model and generate protein-like adjacency matrices as was done in~\cite{bartoli2008effecta}. It also highlights that the proposed heuristic of $P(s)\approx s^{-1}$ is a good starting point to describe amino acid distance distributions. However, here we managed to derive a more broad approximation for which a power law can be seen as a special case~\ref{eq.Ps}.

Gaining a better understanding of the ensemble of folded protein structures can help guide the way to a better understanding of the constraints within which structures may occur. Together with an understanding of secondary structure principles, such as those in~\cite{Danielsson2010, Molkenthin2011}, this can help to narrow down the complex energy landscapes and find paths through them more effectively in the future.

\section*{Data availability statement}
All data generated or analyzed during this study are included in this published article. All raw data and their analysis can be found on GitHub at~\cite{2022sequence}, including information required for their reproduction. 

\section*{Conflict of interests}
The authors declare no competing interests.
\section*{Acknowledgements}
We thank Marc Timme and Matteo Degiacomi for fruitful discussions.
\section*{References}
\bibliographystyle{unsrt}
\bibliography{main}

\end{document}